\documentclass[onecolumn,showpacs,preprintnumbers,amsmath,amssymb,aps,pra]{revtex4}
\usepackage{mathrsfs}
\usepackage{txfonts}
\usepackage{amsmath}
\usepackage{graphicx}
\usepackage{bm}
\usepackage{ulem}

\begin{document}

\title{Revisit the Poynting vector in $\mathcal{PT}$-symmetric coupled waveguides}

\author{Xin-Zhe Zhang, Ru-Zhi Luo, and Jing Chen} \email{jchen4@nankai.edu.cn}

\address{$^1$ MOE Key Laboratory of Weak-Light Nonlinear Photonics, School of Physics, Nankai University, Tianjin 300071, China \\
$^2$ Collaborative Innovation Center of Extreme Optics, Shanxi University, Taiyuan, Shanxi 030006, China
}

\date{\today}

\begin{abstract}
We show that the time-averaged Poynting vector of $\vec{S}=\vec{E}\times \vec{H}^*/2$  in parity-time ($\mathcal{PT}$) symmetric coupled waveguides is always positive and cannot explain the stopped light at exceptional points (EPs). In order to solve this paradox, we must accept the fact that the fields $\vec{E}$ and $\vec{H}$ and the Poynting vector in non-Hermitian systems are in general complex. Based on the original definition of the instantaneous Poynting vector $\vec{S}=\vec{E}\times \vec{H}$, a formula on the group velocity is proposed, which agrees perfectly well with that calculated directly from the dispersion curves. It explains not only the stopped light at EPs, but also the fast-light effect near it. This investigation bridges a gap between the classic electrodynamics and the non-Hermitian physics, and highlights the novelty of non-Hermitian optics.
\end{abstract}

\maketitle

\section{Introduction}

In the past decades we have witnessed the rapid advances of non-Hermitian physics, including the parity-time ($\mathcal{PT}$) symmetry and the exceptional points (EPs) \cite{R01, R02, R03, R04, R05, R06, R07, R08, R09, R10, R11, R12, R13, R14, R15, R16, R17, R18, R19, R20}. By taking into account the active role of loss and gain, non-Hermitian physics applies to open systems without the energy-conservation requirement, and provides a much broader platform for scientists in revealing new physics and exotic applications which have no counterpart in common Hermitian physics. Notable achievements of $\mathcal{PT}$-symmetric optics include the unidirectional invisibility, enhanced sensors, chiral manipulation and various vortex lasers \cite{R01, R02, R03, R04, R05, R06, R07, R08, R09, R10, R11, R12, R13, R14, R15, R16, R17, R18, R19, R20}.

Dynamics of optical waves in $\mathcal{PT}$ symmetric optical systems, e.g. lattices and coupled waveguides, have also been investigated \cite{R21, R22, R23, R24, R25, R26, R27, R28, R29, R30}. For example, Markis \textit{et. al.} showed that $\mathcal{PT}$ periodic structures can exhibit unique characteristics such as double refraction, power oscillations, and nonreciprocal diffraction patterns, which stem from the non-orthogonality of the associated Floquet-Bloch modes \cite{R21}.  Longhi studied the nonreciprocal Bloch oscillations in complex lattices with $\mathcal{PT}$ symmetry \cite{R22}. Stopped light at EP in $\mathcal{PT}$-symmetric waveguides and the associated topological applications have been discussed by various groups \cite{R13, R14, R23}.

Indeed, the great advances of the $\mathcal{PT}$-symmetric optics expand the scope of the research based on the optical-quantum analogies \cite{R31}, especially given that the non-Hermitian quantum physics \cite{R32} is now replacing the Hermitian one. However, this new discipline is still in its infancy. For some special phenomena and effects in the $\mathcal{PT}$-symmetric optics, by judiciously examining details of the different approaches based on the non-Hermitian physics and the classic electrodynamics \cite{R33, R34}, it is inevitable to find some paradoxes that need to be solved \cite{R03, R08}.

Here, we would like to point out and solve a paradox in the stopped-light effect in $\mathcal{PT}$-symmetric coupled optical waveguides \cite{R13}. We analyze the optical fields at EP by using Maxwell's equations, and show that the time-averaged flux of energy defined by the real part of the Poynting vector $\vec{S}^{\{0\}}=\vec{E}\times \vec{H}^*/2$ is always positive. It cannot explain the stopped light at EPs. To solve this paradox, we argue that the general belief, that the fields $\vec{E}$ and $\vec{H}$ are real, must be abandoned when studying the dynamics of optical waves in $\mathcal{PT}$-symmetric systems. In principle, the Poynting vector is instantaneous complex and time-space-dependent in any non-Hermitian optical system, and is in the original form of $\vec{S}^{\{2\omega\}}=\vec{E}\times \vec{H}$. Based on this argument, a simple formula of the group velocity is proposed, which agrees well with that directly calculated from the dispersion curves. This formula explains not only the stopped light at EP, but also the fast-light effect near it. This study bridges a gap between the classic electrodynamics and the non-Hermitian physics, highlights the novelty of $\mathcal{PT}$ symmetry, and contributes to the advances of non-Hermitian optics and slow/fast-light science.

The structure of this article is organized as follows. In subsection 2.1, we analyze the eigenmodes of optical waves in a $\mathcal{PT}$-symmetric coupled waveguides by using the rigorous Maxwell's equations. We calculate the dispersion curves of the eigenmodes and show that it repeats the standard results of $\mathcal{PT}$ symmetry and supports an EP. We also show that the group velocity calculated from the dispersion curves indeed supports stopped light at EP. In subsection 2.2, we emphasize that there is a paradox in the group velocity calculated with the time-averaged complex Poynting vector $\vec{S}^{\{0\}}$. The time-averaged flux of energy is always positive (it is null only when $E$ or $H$ is zero) and the associated group velocity cannot be zero. In subsection 2.3, we argue that in order to solve this paradox we must accept the fact that all the field components of an optical wave are generally complex, and utilize the instantaneous Poynting vector $\vec{S}^{\{2\omega\}}$ instead. Furthermore, a general formula about the group velocity is proposed. By using this formula we explain that the stopped light at EP and the fast-light effect near it are both non-Hermitian interference effects that rely on the instantaneous spatially-distributed complex energy flux and energy density. Discussion about the deep-lying physics and potential challenges is provided at the end of this article.

\begin{figure}
\centerline{\includegraphics[width=9cm]{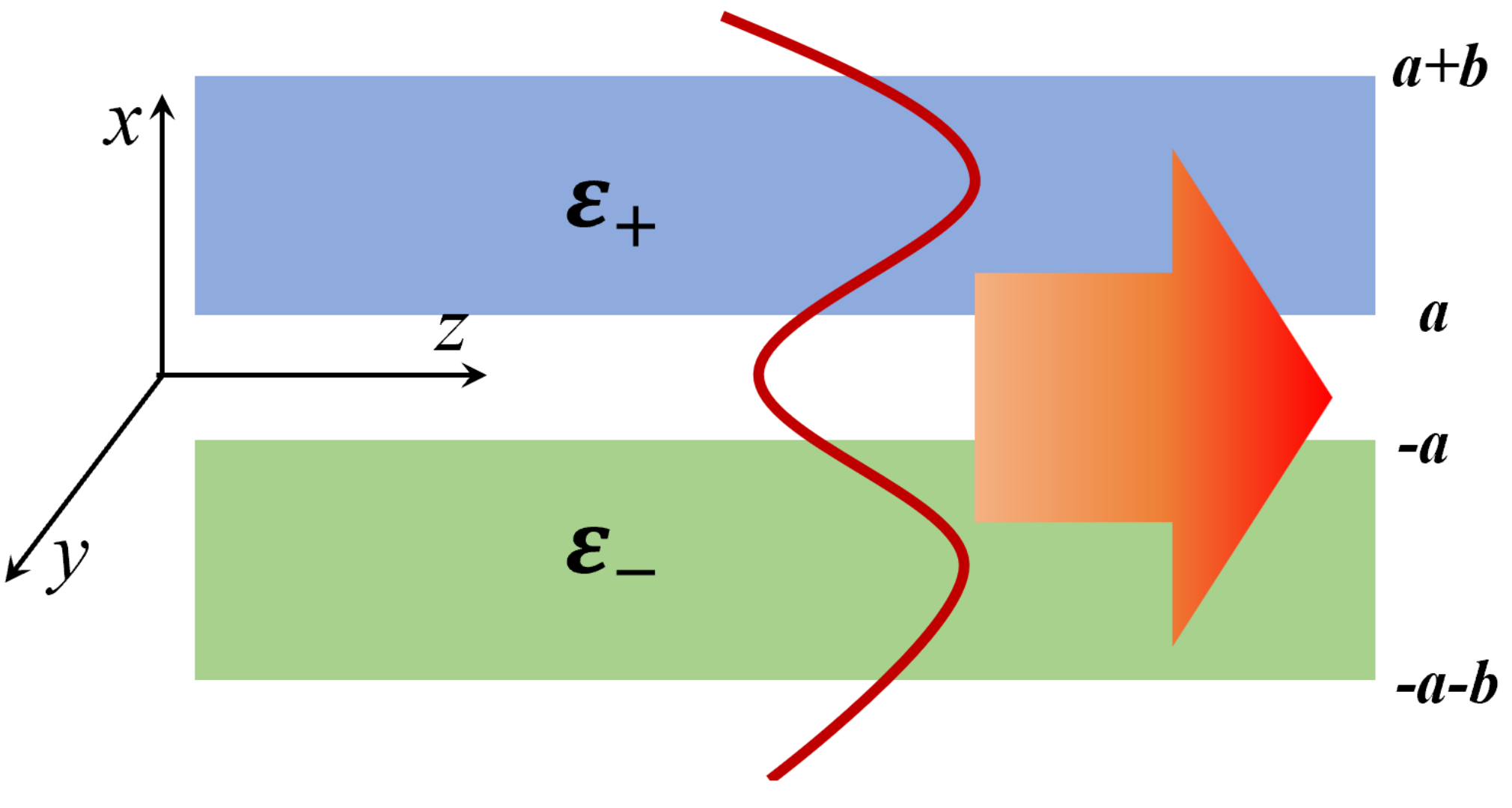}} \caption{Structure of the investigated $\mathcal{PT}$-symmetric coupled optical waveguides.}
\end{figure}

\section{Structure and analysis}

\subsection{Structure and analysis by using Maxwell's equations}

The structure under investigation is shown in Fig. 1. It is a simple optical system with two straight parallel waveguides. The two waveguides are geometrically identical and infinite in the $y-z$ plane. Width of each waveguide is $b=400$ nm, and distance between them is $2a=300$ nm.  The dielectric constants of them are conjugate to each other, that $\epsilon_\pm=(n_r\pm jn_i)^2=(1.5\pm j0.03)^2$. The surrounding medium is vacuum with $\epsilon=1$. The propagating mode inside this structure can be studied by using Maxwell's equations without free carriers and currents. Here, without loss of generality, let us consider the propagation of a transverse electrical (TE) mode in the $z$ direction. The transverse $E_y$ fields can be expressed as
\begin{equation}
E_y= e^{-jkz+j\omega t}\left\{
\begin{array}{cc}
  Ae^{-\beta(x-a-b)}, &x>a+b\\
  B_1e^{-j\alpha_+ (x-a)}+B_2e^{+j\alpha_+ (x-a)}, &a+b>x>a \\
  C_1e^{+\beta x}+C_2e^{-\beta x}, &+a>x>-a \\
  D_1e^{-j\alpha_- (x+a)}+D_2e^{+j\alpha_- (x+a)}, &-a>x>-a-b \\
  Fe^{+\beta(x+a+b)}, &x<-a-b\\
  \end{array}\right.
\end{equation}
where
\begin{equation}
k^2-\beta^2=\omega^2/c^2,
\end{equation}
\begin{equation}
k^2+\alpha_\pm^2=\epsilon_\pm\omega^2/c^2.
\end{equation}
The associated magnetic fields can be found from $\nabla\times \vec{E}=-\partial \vec{B}/\partial t$, which gives
\begin{equation}
H_x=-\frac{k}{\mu_0\omega}E_y,
\end{equation}
\begin{equation}
H_z= \frac{j}{\mu_0\omega}e^{-jkz+j\omega t}\left\{
\begin{array}{cc}
  -\beta Ae^{-\beta(x-a-b)}, &x>a+b\\
  -j\alpha_+ [B_1e^{-j\alpha_+ (x-a)}-B_2e^{+j\alpha_+ (x-a)}], &a+b>x>a \\
  \beta [C_1e^{+\beta x}- C_2e^{-\beta x}], &+a>x>-a \\
  -j\alpha_-[D_1e^{-j\alpha_- (x+a)}-D_2e^{+j\alpha_- (x+a)}], &-a>x>-a-b \\
  \beta Fe^{+\beta(x+a+b)}, &x<-a-b\\
  \end{array}\right.
\end{equation}

Dispersion of the eigenmode can be found from above expressions by applying electromagnetic boundary conditions. Defining
\begin{equation}
F_\pm=\frac{\beta+j\alpha_\pm}{\beta-j\alpha_\pm},
\end{equation}
\begin{equation}
\Upsilon_\pm=\frac{F_\pm \exp(j2\alpha_\pm b)-F_\pm^{-1}}{\exp(j2\alpha_\pm b)-1},
\end{equation}
the dispersion is governed by
\begin{equation}
\Upsilon_+\Upsilon_--\exp(-4\beta a)=0.
\end{equation}

By using Eq. (8) we calculate the dispersion of the TE eigenmodes. As shown in Fig. 2(a), generally we can achieve two branches of eigenmodes when $k$ is relatively small. When $k$ increases further an EP can be accessed, where the two branches of dispersion coalesce. By differentiating the dispersion in Fig. 2(a) we can find the associated group velocity
\begin{equation}
v_g^{\{D\}}=\frac{\partial\omega}{\partial k},
\end{equation}
where the superscript $'D'$ implies that it is calculated from the dispersion curve. From the result shown in Fig. 2(b) we can see near EP one branch of $v_g^{\{D\}}$ drops to zero directly, while the other branch of $v_g^{\{D\}}$  increases sharply toward infinite. Figure 2 confirms the conclusions made in \cite{R13}, that the $\mathcal{PT}$-symmetric coupled optical waveguides support stopped light of $v_g^{\{D\}}=0$ at EP and fast light of $v_g^{\{D\}}>c$ near it.

\begin{figure}
\centerline{\includegraphics[width=13cm]{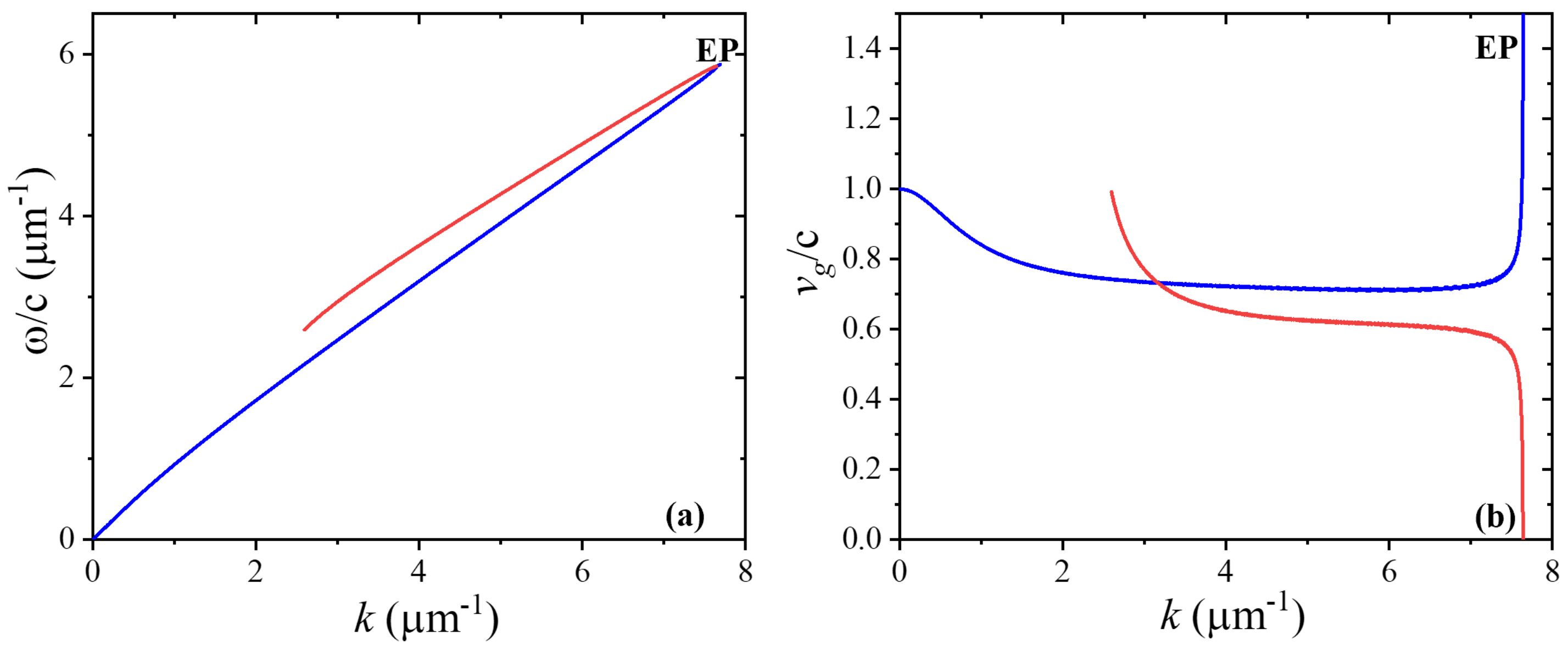}} \caption{(a) Dispersion of the guided TE modes, and (b) the associated group velocity $v_g^{\{D\}}$ versus wavevector $k$. }
\end{figure}

\subsection{Paradox about the Poynting vector and the energy flux}

Above analysis proves that Maxwell's equations and boundary conditions can be applied to the investigation of $\mathcal{PT}$-symmetric optical systems. It confirms most features of the $\mathcal{PT}$ symmetry especially the emergence of EP. Because Maxwell's equations are  rigorous and do not contain any prerequisite \cite{R33, R34}, e.g. all the parameters inside can be complex, they can describe the field dynamics in all kinds of non-Hermitian optical systems. The results from Maxwell's equations then can be utilized as the reference or accurate answers in criticizing other theories and models, as done in this article.

Recently it was shown that EPs stop light \cite{R13}. Accounting to the classic electrodynamics \cite{R33, R34, R35}, the group velocity $v_g$ of the guided mode can also be found by using the time-averaged flux of energy given by the $z$-directional Poynting vector $S^{\{0\}}_z$ and the time-averaged energy density $W^{\{0\}}$ (page 364 of \cite{R33}), as
\begin{equation}
v_g^{\{0\}}=\frac{S_z^{\{0\}}}{W^{\{0\}}},
\end{equation}
where the superscript $'0'$ implies that the parameter is a time-independent constant, and
\begin{equation}
S_z^{\{0\}}=\int s_z^{\{0\}} dx=\int \frac{1}{2}(\vec{E}\times \vec{H}^*)_z dx,
\end{equation}
\begin{equation}
W^{\{0\}}=\int\frac{1}{4}\{ \frac{\partial(\varepsilon\omega)}{\partial \omega}|E|^2+\frac{\partial(\mu\omega)}{\partial \omega}|H|^2\}dx.
\end{equation}
The time-averaged energy density $W^{\{0\}}$ is closely related to the dispersions in $\varepsilon$ and $\mu$. However, one of the main conclusions in \cite{R13} is that the stopped light is independent of the material dispersion and is solely based on the $c$-product self-orthogonality of the eigenmode \cite{R32}. Also dispersions in $\varepsilon$ and $\mu$  are neglected here when calculating the results shown in Fig. 2. Consequently, in order to explain the stopped light we only need to consider the condition of $S_z^{\{0\}}=0$. The case of divergent $W^{\{0\}}$, which can be achieved at a high-DOS point near an intrinsic material resonance such as that of the Lorentz dispersion, does not need to be considered.

\begin{figure}
\centerline{\includegraphics[width=9cm]{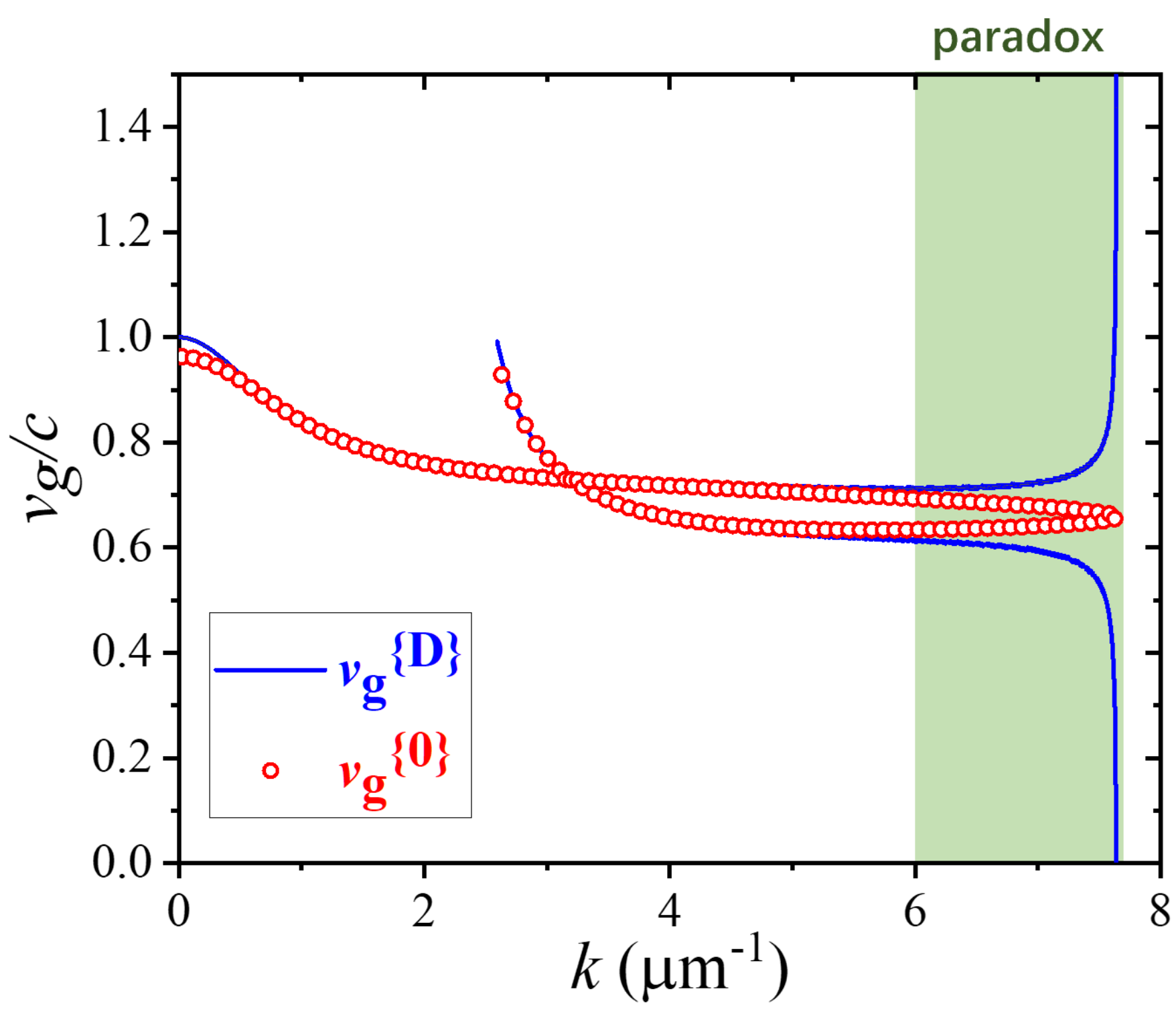}} \caption{The variations of group velocities $v_g^{\{D\}}$ (blue lines) and $v_g^{\{0\}}$ (red circles) versus wavevector $k$.  Near EP the value of $v_g^{\{0\}}$ satisfies $c>v_g^{\{0\}}>0$, and there exists a paradox between $v_g^{\{D\}}$ and $v_g^{\{0\}}$. }
\end{figure}

The value of $S_z^{\{0\}}$ can be obtained by the integral of $s_z^{\{0\}}$ following Eq. (11). However, substituting Eq. (4) into this formula we meet a paradox of
\begin{equation}
s_z^{\{0\}}= -\frac{1}{2}E_yH_x^*=\frac{k^*}{2\mu_0\omega^*} E_y E_y^*>0,
\end{equation}
that the direction of the energy flux is positive at any spatial position $x$ where $E_y$ is not null.  Such a conclusion applies not only to EP, but also to the exact $\mathcal{PT}$ phase before EP, where two dispersion branches exist. If we consider the transverse magnetic mode, although the expression of $s_z^{\{0\}}$ changes to $s_z^{\{0\}}= k^* H_y H_y^*/(2\epsilon^*\omega^*)$, the real part of $\epsilon$ in the whole $\mathcal{PT}$ symmetric structure is positive, so $s_z^{\{0\}}$ is still positive. The imaginary part of $\epsilon$ does not deny this conclusion because $n_i$ is much smaller than $n_r$.

To provide a clear demonstration of above paradox, we calculate the variation of $v_g^{\{0\}}$ versus $k$ by employing Eq. (10), and show it in Fig. 3. For comparison, the results of $v_g^{\{D\}}$ shown in Fig. 2(b) are copied here. We can see in the long-wavelength regime these two approaches agree well (the weak deviation near zero $k$ is due to the limited spatial size in $x$ when performing the integral). But when $k$ increases and the $\mathcal{PT}$ phase approaches the colascent phase transition point of EP, these two approaches diverge and no longer fit (see the shadow region in Fig. 3). The value of $v_g^{\{0\}}$ is always positive, and does not drop to zero. It is also always smaller than $c$, and does not support fast light. Since the results shown in Fig. 2 are obtained from the rigorous Maxwell's equations, it is obvious that Eq. (10) must be examined very carefully in order to solve the paradox of
\begin{equation}
c>v_g^{\{0\}}>0.
\end{equation}

\subsection{The fields are complex in non-Hermitian systems}

To solve this paradox, we must return to the initial thought about the Poynting vector. In fact, the initial definition of Poynting vector comes directly from Maxwell's equations \cite{R33, R34, R35}, which gives
\begin{equation}
\vec{s}^{\{2\omega\}}= \vec{E}\times \vec{H}.
\end{equation}
Since $\vec{E}$ and $\vec{H}$ are harmonic oscillating fields with time- and space-dependent terms of $\exp(-jkz+j\omega t)$, $\vec{s}^{\{2\omega\}}$ is the instantaneous complex Poynting vector that oscillates periodically in space and time following $\exp(-j2kz+j2\omega t)$, and its magnitude is sensitive to the phases of $\vec{E}$ and $\vec{H}$. It is why we utilize the superscribe $'2\omega'$ here in order to distinguish it from the time-averaged one $s_z^{\{0\}}$.

In the classic electrodynamics \cite{R33, R34, R35}, the transformation from $\vec{s}^{\{2\omega\}}$ to $\vec{s}^{\{0\}}$ is based on a well-known assumption that all the fields $\vec{E}$ and $\vec{H}$ in the Maxwell's equations are real so as to be experimentally measurable. Evidently, this assumption is the key to solving the above paradox. Here we would like to argue that such an assumption must be abandoned when applied to the non-Hermitian systems because the fields should, in general, be complex. The reasons are as follows.

Firstly,  in Maxwell's equations the $\vec{E}$ and $\vec{H}$ fields are connected by $\nabla\times \vec{H}=\partial \vec{D}/\partial t$. Because $\epsilon$ in a non-Hermitian system is complex, the simultaneous requirement of real valued $\vec{E}$ and $\vec{H}$ cannot be satisfied.

Secondly, and the most important, each single field in the coupled $\mathcal{PT}$-symmetric waveguides suffers from a phase-asynchronous mechanism in the exact $\mathcal{PT}$ phase and EP. To be more explicit, choosing either $\vec{E}$ or $\vec{H}$ as the investigated physical quantity, albeit in the two waveguides they have the same magnitude, their phase difference is not exact $0$ or $\pi$. Considering the extreme scenario at EP, the phase difference in $E$ inside the two waveguides is $\pi/2$. The requirement of real $E$ then renders that at some instantaneous moments the fields $\text{Re}\{E\}$ in the two waveguides are not equal. For example, when $\text{Re}\{E\}$ in one waveguide reaches the maximum value, in the other waveguide $\text{Re}\{E\}$ should be zero. Such an operation, in fact, transforms the self-orthogonalized singular eigenstate of $[E_1, E_2]^T=[1,i]^T$ (or $[1,-i]^T$) at EP, where $E_{1,2}$ represents the field component in one constituent waveguide, into the $[1,0]^T$ or $[0,1]^T$ eigenstate. As a result,  the singularity at EP is broken, and the $\mathcal{PT}$-symmetric optical system is converted into a Hermitian one. Such a process is not acceptable when studying $\mathcal{PT}$ symmetry, and, evidently, the requirement of real fields may not apply to non-Hermitian systems. Figure 3 also supports our consideration because $v_g^{\{D\}}$ and $v_g^{\{0\}}$ diverge extremely strong near EP (see the paradox region in Fig. 3). Note that the paradox is evident only near EP (see Fig. 3). In order to prove the feasibility of above analysis the whole experimental set-up should be operated exactly at EP, which is a challenging task because EPs are very sensitive to perturbation \cite{R10, R11, R12, R25}.

\begin{figure}
\centerline{\includegraphics[width=14cm]{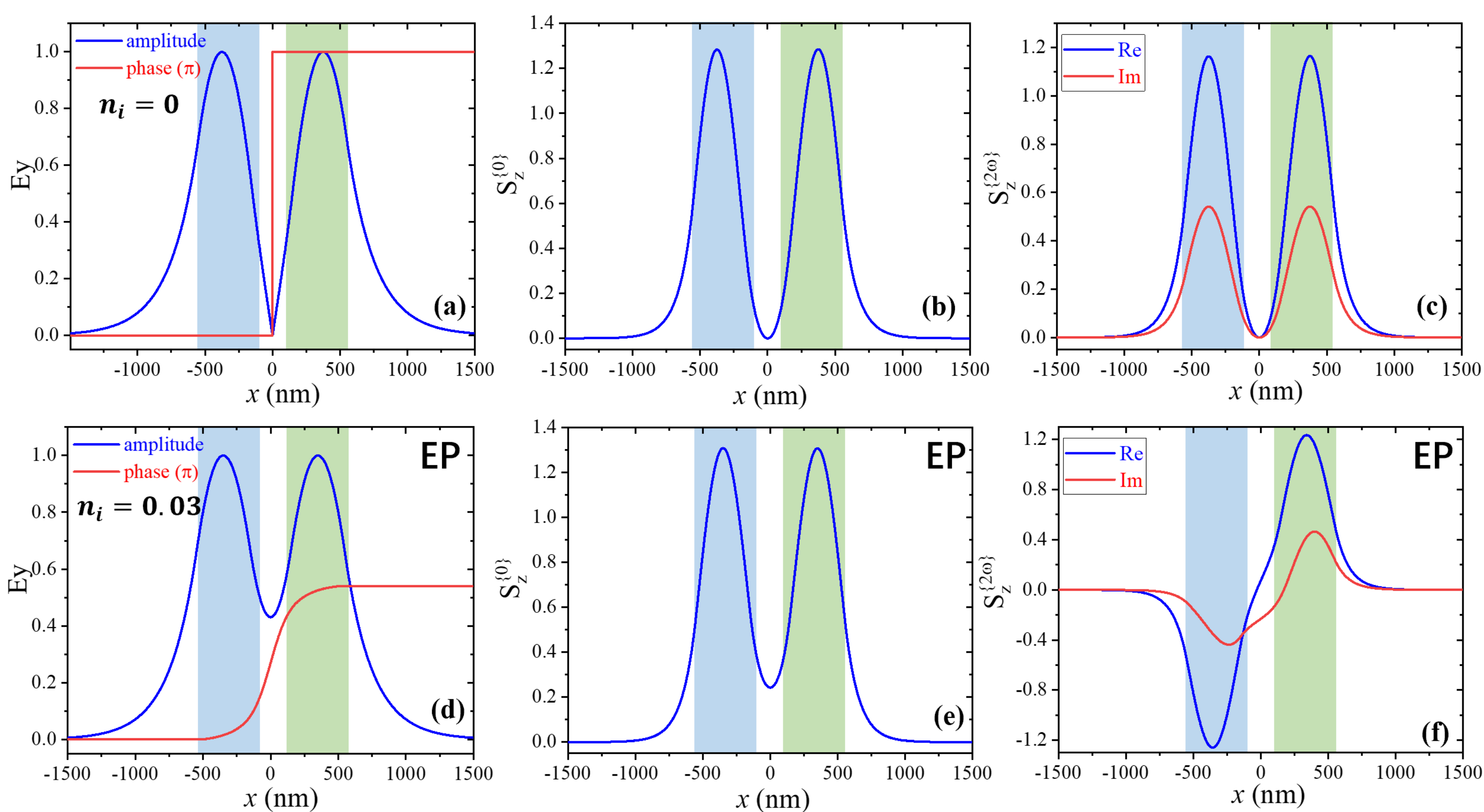}} \caption{Distributions of (a,d) the amplitude and phase of $E_y$, (b,e) the time-averaged $s^{\{0\}}_z$, and (c,f) the instantaneous complex $s^{\{2\omega\}}_z$ for (a-c) the antisymmetric mode when $n_i=0$, and (e-f) the coalescent mode at EP when $n_i=0.03$, respectively.}
\end{figure}

To provide more evidences about above discussion we calculate the distributions of the complex field $E_y$, $s^{\{0\}}_z$ and $s^{\{2\omega\}}_z$ at an anti-symmetric mode when $n_i=0$ (Hermitian), and at EP when $n_i=0.03$ (non-Hermitian). These two scenarios have the same wavevector but slightly different $\omega$ values. Furthermore, an additional phase is introduced to the fields to check the influence over the Poynting vector. From the results shown in Fig. 4 we can see in both cases the time-averaged $s^{\{0\}}_z$ is positive [Figs. 4(b) and 4(d)], in agreement with Eq. (13). As for the instantaneous Poynting vector $s^{\{2\omega\}}_z$, in the Hermitian case of $n_i=0$ the real and imaginary parts of it in the whole structure are of the same phase [Fig. 4(c)]. On the contrary, at EP $s^{\{2\omega\}}_z$ can have opposite instantaneous directions in the whole $\mathcal{PT}$ structure, see Fig. 4(f). The strict requirement of real $\vec{E}$ and $\vec{H}$ would forbid such a kind of opposite-directional $s^{\{2\omega\}}_z$ shown in Fig. 4(f) because this phenomenon asks for a $\pi/2$ phase difference in the $E$ fields [see Fig. 4(d)].

Figure 4(f) prompts us that the spatial-averaged instantaneous complex Poynting vector $\vec{s}^{\{2\omega\}}$ might be zero at EP, and might explain the stopped light. To check it, we calculate the $z$-directional component of it by using
\begin{equation}
S_z^{\{2\omega\}}=\int s_z^{\{2\omega\}} dx,
\end{equation}
in the whole dispersion curves of Fig. 2(a). However, to provide a comparable analysis the field should be properly normalized. Since $s_z^{\{2\omega\}}$ has a time- and space-dependence of $\exp(-j2kz+j2\omega t)$, the normalization should be based on a quadratic combination of  $\vec{E}$, $\vec{D}$, $\vec{H}$, and $\vec{B}$.

By analyzing the calculated fields from Maxwell's equations in subsection 2.1, we find two equations for the configuration studied here. The first equation is that at any instantaneous moment,
\begin{equation}
W^{\{2\omega\}}=\int D_yE_ydx=\int (B_xH_x-B_zH_z)dx.
\end{equation}
This equation is equivalent to the well-known classic one that the time-averaged field power is equally distributed in the electric and magnetic components, that
$\int\vec{D}\cdot\vec{E^*}dx=\int \vec{B}\cdot\vec{H^*}dx$. Note that if intrinsic dispersion \cite{R03} is taken into account, a more complex expression need to be found \cite{R28, R29, R30}.

\begin{figure}
\centerline{\includegraphics[width=9cm]{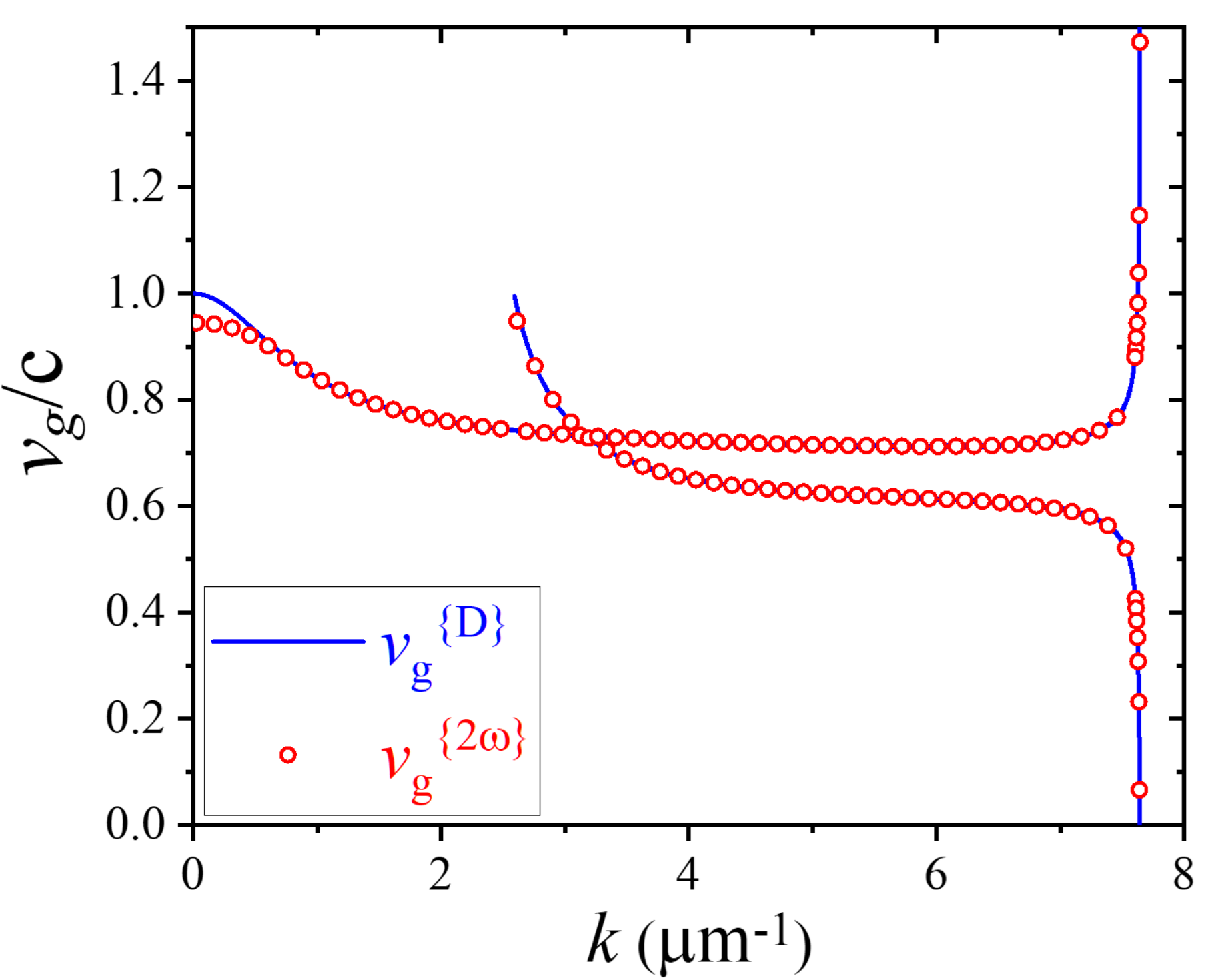}} \caption{The variations of the group velocities $v_g^{\{2\omega\}}$ (red circles) and $v_g^{\{D\}}$ (blue lines) versus wavevector $k$. }
\end{figure}

The second equation is that the group velocity can be defined by
\begin{equation}
v_g^{\{2\omega\}}=\frac{S_z^{\{2\omega\}}}{W^{\{2\omega\}}},
\end{equation}
This equation is similar to Eq. (10), but because the field is not required to be real, now the parameters $S_z^{\{0\}}$ and $W^{\{0\}}$ are replaced by the instantaneous complex parameters $S_z^{\{2\omega\}}$ and $W^{\{2\omega\}}$, respectively.

We calculate the value of $v_g^{\{2\omega\}}$ versus $k$, and find that it is always real and agrees well with $v_g^{\{D\}}$, see Fig. 5. Consequently, it can explain two puzzles in $v_g^{\{D\}}$ shown in Fig. 2(b). The first puzzle is about the zero group velocity at EP, which, as expected from above analysis, is rooted at the zero instantaneous complex Poynting vector, that
\begin{equation}
S_z^{\{2\omega\}}=0 \longrightarrow v_g^{\{D\}}=0.
\end{equation}
The second puzzle is about the fast light, where one branch of the group velocity is greater than $c$ and approaches infinite. This fast-light effect can be explained from the complex instantaneous energy density $W^{\{2\omega\}}$, that
\begin{equation}
W^{\{2\omega\}}=0 \longrightarrow v_g^{\{D\}}=\infty.
\end{equation}
Because the electric fields in the two waveguides are of a phase difference near $\pi/2$, the weighted integral of $\epsilon \vec{E}^2$ in the whole structure, where $\epsilon$ is equivalent to the weight, can approach zero. Clearly, such a zero-$W^{\{2\omega\}}$ effect cannot be achieved if we assume that all fields are real in optical waves.

\section{Discussion}

To this end, we can use Eq. (18) to explain all the features on the dispersion curves of the $\mathcal{PT}$ symmetric waveguides, not only the stopped light at EP but also the fast light. Albeit the definition of the complex instantaneous energy density $W^{\{2\omega\}}$ has not taken into account the influence of the material dispersion \cite{R03}, and the trivial form of $W^{\{2\omega\}}$ in Eq. (17) might be an occasional one and could not be applied to other structural geometries or other polarized modes, our analysis presented above clearly proves that we must abandon the belief that all fields are real in electrodynamics, and the energy flux defined by the Poynting vector should be applied very carefully, especially in non-Hermitian systems. It is just the goal of the present article. In the successive research we can try to reveal the deep-lying physics of these trivial expressions of $S_z^{\{2\omega\}}$ and $W^{\{2\omega\}}$, and develop the exact forms of them when dispersions in $n_r$ and $n_i$ are taken into account \cite{R03, R36, R37}.

In fact, it is not an astonishment that the definition of $S_z^{\{2\omega\}}$ is the correct one for optical field in non-Hermitian systems. We can explain it from another perspective by analogizing between optics and quantum theory. In Hermitian systems, the orthonormal property requires that each eigenfunction $\varphi_i$ satisfies the scalar-product of $\int \varphi_i^*\varphi_jdx=\delta_{ij}$. However, in non-Hermitian systems the $c$-product should be used \cite{R13, R32}, that  $\int \varphi_i\varphi_jdx=\delta_{ij}$. Based on the optical-quantum analogue, it would not be a surprise that the seemly correlated definitions of the scalar-product and the Poynting vector in the Hermitian frame,
\begin{equation}
\int \varphi_i^*\varphi_jdx=\delta_{ij}\longrightarrow \vec{S}=\vec{E}\times \vec{H}^*,
\end{equation}
where a conjugate operation is required, can be transformed to
\begin{equation}
\int \varphi_i\varphi_jdx=\delta_{ij}\longrightarrow \vec{S}=\vec{E}\times \vec{H}
\end{equation}
in the non-Hermitian frame. As for $W^{\{2\omega\}}=\int \epsilon \vec{E}^2 dr$, it is just the analogue of the normalization factor on the non-Hermitian quantum state $\vec{E}$, with an additional weight of $\epsilon$ \cite{R32}. Such a topic deserves our further discussion.

The importance of this work is that we have proved that in non-Hermitian optics all the fields are generally complex. The stopped light at EP and fast light near it are interference phenomena because they cannot be obtained in a single waveguide. They rely on the overall integral of the quantity $S_z^{\{2\omega\}}$ (also $W^{\{2\omega\}}$), which have different complex values at different spatial positions so that they can cancel each other out when performing the integral. Such an effect has no counterpart in Hermitian systems \cite{R36, R37}. The complex nature of fields in non-Hermitian systems has not been noticed before because, to our belief, their distinct signature can be noticed only near EPs, see Fig. 2. Away from EPs the analysis by using the time-averaged $\vec{S}^{\{0\}}$ and $W^{\{0\}}$ still gives tolerable results.

Before ending this article, we would like to provide advanced comments on the potential investigation targets, especially the experimental ones. Firstly, the instantaneous parameters $S_z^{\{2\omega\}}$ and $W^{\{2\omega\}}$ contain factor of $\exp(j2\omega t)$, and the time-averaged process would wash them out. Then it is a challenge to answer what a parameter we should (or could) observe in optical experiments if the field is complex, and whether the stopped light at EP can be readily observed in experiments. We check our simulation, and find that
\begin{equation}
\frac{S_z^{\{2\omega\}}}{W^{\{2\omega\}}}=\frac{\text{Re}\{S_z^{\{2\omega\}}\}}{\text{Re}\{W^{\{2\omega\}}\}}
\end{equation}
holds true at any instantaneous moment. Consequently, we can still say that the energy density and the flux of energy in non-Hermitian optical systems are real, and they are given by the real part of $W^{\{2\omega\}}$ and $S_z^{\{2\omega\}}$, respectively.  However, even in this scenario we should accept the fact that $s_z^{\{2\omega\}}=\text{Re}\{\vec{E}\times \vec{H}\}$ and the local density of energy $w^{\{2\omega\}}=\text{Re}\{\epsilon \vec{E}^2\}$ could be negative in order to explain $v_g^{\{2\omega\}}=0$ and $v_g^{\{2\omega\}}>c$. Equation (23) only implies that $W^{\{2\omega\}}$ and $S_z^{\{2\omega\}}$ are synchronized together and governed by the group velocity.

Secondly, a successful observation of the stopped light at EP would support our conclusion that any electromagnetic fields in nature are complex and no longer real, and some conclusions in the textbooks of classic electrodynamics should be checked again. If our former belief, that all the field $\vec{E}$ and $\vec{H}$ are real, is still correct, then the stopped light at EP cannot be observed. Former literatures about EPs in straight waveguides should also be checked very carefully. Any efforts in rising to this challenge will contribute to the advances of the science and engineering of optics and optical-non-Hermitian analogues.

\section{Conclusion}

In summary, here we show that the time-averaged Poynting vector of $\vec{S}^{\{0\}}=\vec{E}\times \vec{H}^*/2$  in a $\mathcal{PT}$-symmetric coupled waveguide is always positive and cannot explain the stopped light at EP. In order to solve this paradox, we must accept the fact that the field components $\vec{E}$ and $\vec{H}$ and the Poynting vector in non-Hermitian systems are complex. Based on the original definition of the Poynting vector $\vec{S}^{\{2\omega\}}=\vec{E}\times \vec{H}$, a formula on the group velocity is proposed, which is in agreement with that directly calculated from the dispersion curves. It explains that both the stopped light at EP and fast light near it are non-Hermitian interference effects that rely on the instantaneous spatially-distributed complex energy flux and energy density. This study bridges a gap between the classic electrodynamics and the non-Hermitian physics, highlights the novelty of  $\mathcal{PT}$ symmetry, and contributes to the advances of non-Hermitian optics and slow/fast-light science.

\section*{Funding}
Natural National Science Foundation of China (NSFC) (11874228, 12274241).

\section*{Disclosures}

The authors declare no conflicts of interest.

\section*{Data availability}
Data underlying the results presented in this paper are not publicly available at this time but may
be obtained from the authors upon reasonable request.


\end{document}